 \documentclass[reprint, amsmath, amssymb, aps, prb]{revtex4-1}
\usepackage[]{graphicx}      % Include figure files
\usepackage{bm}             	% bold math
\usepackage{times}			% Times font
\usepackage{tabularx}
\usepackage{color}
\usepackage{dcolumn}		% Align table columns on decimal point
\usepackage{amsmath}
\usepackage{amssymb}
\usepackage{amsfonts}
\usepackage{times}
\usepackage{tabularx}
\usepackage{xcolor,colortbl}
\usepackage[T1]{fontenc}
\usepackage[utf8]{inputenc}

\begin{document}
\title{Gilbert damping of high anisotropy Co/Pt multilayers}
\author{Thibaut Devolder}
\email{thibaut.devolder@u-psud.fr}
\affiliation{Centre de Nanosciences et de Nanotechnologies, CNRS, Univ. Paris-Sud, Universit\'e Paris-Saclay, C2N-Orsay, 91405 Orsay cedex, France}
\author{S. Couet}
\affiliation{imec, Kapeldreef 75, 3001 Heverlee, Belgium}
\author{J. Swerts}
\affiliation{imec, Kapeldreef 75, 3001 Heverlee, Belgium}
\author{G. S. Kar}
\affiliation{imec, Kapeldreef 75, 3001 Heverlee, Belgium}

\date{\today}                                           
%%%%%%%%%%%%%%%%%%%%%%%%%%%%%%%%%%%%%%%%
%
%       Abstract
%
%%%%%%%%%%%%%%%%%%%%%%%%%%%%%%%%%%%%%%%%
\begin{abstract}
Using broadband ferromagnetic resonance, we measure the damping parameter of [Co(5 \r{A})/Pt(3 \r{A})]${\times 6}$ multilayers whose growth was optimized to maximize the perpendicular anisotropy. Structural characterizations indicate abrupt interfaces essentially free of intermixing despite the miscible character of Co and Pt. Gilbert damping parameters as low as 0.021 can be obtained despite a magneto-crystalline anisotropy as large as $10^6~\textrm{J/m}^3$. The inhomogeneous broadening accounts for part of the ferromagnetic resonance linewidth, indicating some structural disorder leading to a equivalent 20 mT of inhomogenity of the effective field. The unexpectedly relatively low damping factor indicates that the presence of the Pt heavy metal within the multilayer may not be detrimental to the damping provided that intermixing is avoided at the Co/Pt interfaces. 
\end{abstract}

%%%%%%%%%%%%%%%%%%%%%%%%%%%%%%%%%%%%%%%%
%
%       Social media Abstract
%
% The Gilbert damping of high anisotropy Co/Pt multilayers can be as low as 0.021.
%
%%%%%%%%%%%%%%%%%%%%%%%%%%%%%%%%%%%%%%%%

%%%%%%%%%%%%%%%%%%%%%%%%%%%%%%%%%%%%%%%%
%
%       Keywords
%
%  Gilbert damping, perpendicular magnetic anisotropy, Co/Pt multilayer.
%
%%%%%%%%%%%%%%%%%%%%%%%%%%%%%%%%%%%%%%%%

\maketitle

%%%%%%%%%%%%%%%%%%%%%%%%%%%%%%%%%%%%%%%%
%
%                Paper
%
%%%%%%%%%%%%%%%%%%%%%%%%%%%%%%%%%%%%%%%%

%%%%%%%%%%%%%%%%%%%%%%%%%%%%%%%%%%
\section{Introduction}
Thanks to their large perpendicular magnetic anisotropy, their confortable magneto-optical signals and their easy growth by physical vapor deposition \cite{mathet_morphology_2003}, the [Co/Pt] multilayers are one of the most popular system in spintronics. Early in spintronics history this model system was used to study the physics of domain wall propagation \cite{lemerle_domain_1998}, for the development of advanced patterning techniques \cite{chappert_planar_1998} and for the assessment of micromagnetic theories \cite{belliard_stripe_1997}. More recently they have been extensively used as high quality fixed layers in perpendicularly magnetized tunnel junctions, in particular in the most advanced prototypes of spin-transfer-torque magnetic random access memories memories \cite{yakushiji_very_2017}. %The magnetization of such multilayers can easily be adjusted by the Co to Pt content ratio \cite{Fujita_damping_2008}
Despite the widespread use of Co/Pt multilayers, their high frequency properties, and in particular their Gilbert damping parameter remains largely debated with experimental values that can differ by orders of magnitude from \cite{Fujita_damping_2008} 0.02 to 20 times larger \cite{metaxas_creep_2007} \textcolor{black}{and \textcolor{black}{theoretical calculations from circa 0.035 in Co$_{50}$Pt$_{50}$ alloys \cite{drchal_ab_2017} to slightly smaller or substantially larger values in multilayers made of chemically pure layers \cite{barati_gilbert_2017}}}. Direct measurements by conventional ferromagnetic resonance (FMR) are scarce as the high anisotropy of the material pushes the FMR frequencies far above \cite{Fujita_damping_2008} 10 GHz and results in a correlatively low permeability that challenges the sensitivity of commercial FMR instruments \cite{yuan_interfacial_2003}. As a result most of the measurements of the damping of Co/Pt systems were made by all-optical techniques \cite{mizukami_gilbert_2010, barman_ultrafast_2007} in small intervals of applied fields. Unfortunately this technique requires the static magnetization to be tilted away from the out-of-plane axis and this tilt renders difficult the estimation of the contribution of the material disorder to the observed FMR linewidth using the established protocols \cite{mizukami_effect_2002}; this is problematic since in Co-Pt systems there contributions of inhomogeneity line broadening and two-magnon scattering by the structural disorder (roughness,  interdiffusion, granularity,...) are often large \cite{mo_origins_2008, schellekens_determining_2013}.

It is noticeable that past reports on the damping of Co/Pt systems concluded that it ought to be remeasured in samples with atomically flat interfaces \cite{mizukami_gilbert_2010}. Besides, this measurement should be done in out-of-plane applied field since this eases the separation of the Gilbert damping contribution to the linewidth from the contribution of structural disorder \cite{shaw_determination_2012}. In this paper, we measure the damping parameter of [Co(5 \r{A})/Pt(3 \r{A})]$_{\times 6}$ multilayers whose growth was optimized to maximize perpendicular anisotropy anisotropy. The sputter-deposition is performed at an extremely low \cite{musil_low-pressure_1998} Argon pressure in remote plasma conditions which enables very abrupt interfaces that are essentially free of intermixing. We show that in contrast to common thinking, the Gilbert damping parameter of Co/Pt multilayers can be low; its effective value is 0.021 but it still likely \cite{shaw_determination_2012} includes contributions from spin-pumping that our protocol can unfortunately not suppress.

%Heavry damping (0.1 to 0.13) is observed in [Co(0.4)/Pt(0.8)]$_{\times 4}$ \cite{barman_ultrafast_2007}. Substantial dercrease of anisotropy with number of repeats, coming together with an increase of alpha from 0.1 to 0.13.
%\cite{Fujita_damping_2008} [Co(1)/Pt(1.2)]$_{\times 12}$. Q-band FMR (35 GHz); g=2.2, like buck Cobalt. alpha =0.037
%interlayer spin-pumping (like in Fe/Au/Fe) : non existing as the different Co layers are exactly similar (precess in phase) Cap and buffer spin-pumping (like in \cite{shaw_determination_2012})
%\cite{schellekens_determining_2013} "inhomogeneous broadening seems more pronounced for samples with a smaller perpendicular anisotropy"

%%%%%%%%%%%%%%%%%%%%%%%%%%%%%%%%%%%%%%%%%%%%%%%%%%%%%%%%%%%%%%
\section{Experimental}
\begin{figure}
\centering
\includegraphics[width=9 cm]{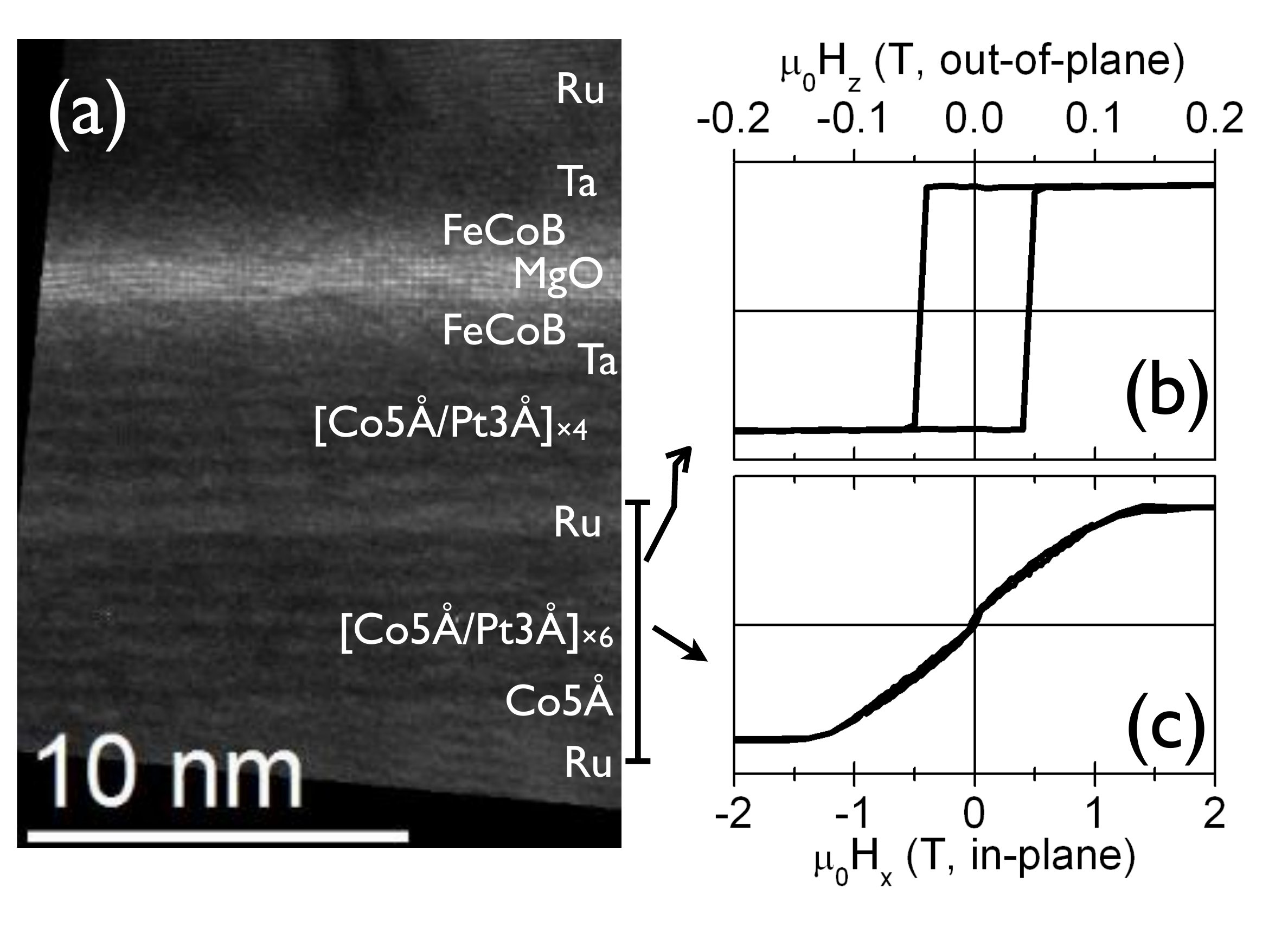}
\caption{(Color online). Structure and anisotropy of a Co-Pt multilayer. (a) Transmission Electron Micrograph of a magnetic tunnel junction that embodies our Co/Pt as hard multilayer at the bottom of the reference synthetic antiferromagnet, similar to that of ref. \onlinecite{devolder_time-resolved_2016}. (b) Easy axis and (c) hard axis hysteresis loops of the hard multilayer \textcolor{black}{when covered with Ru(70 \r{A})/Ta(70 \r{A})/Ru(100 \r{A})/Ta(10 \r{A}, cap)}}
\label{TEM}
\end{figure}

Our objective is to report the high frequency properties of Co/Pt multilayers that were optimized for high anisotropy. The multilayer is grown by sputter-deposition on a Ru (50 \r{A}) buffer and \textcolor{black}{capped with a Ru(70 \r{A})/Ta(70 \r{A})/Ru(100 \r{A})/Ta(10 \r{A}, cap) sequence (bottom to top order)}. \textcolor{black}{The Ru buffer was chosen because it does not mix with Co-based multilayers even under tough annealing conditions \cite{liu_seed_2017}}. The stacks were deposited by physical vapor deposition in a Canon-Anelva EC7800 300 mm system on oxidized silicon substrates at room temperature. The Argon plasma pressure is kept at 0.02 Pa, i.e. substantially lower than the usual conditions of 0.1-0.5 Pa used in typical deposition machines \cite{musil_low-pressure_1998}. As this multilayer is meant to be the reference layer of bottom-pinned magnetic tunnel junctions, in some samples (fig.~\ref{TEM}) the non-magnetic cap is replaced the following sequence: Ta cap /  Fe$_{60}$Co$_{20}$B$_{20}$ / MgO / Fe$_{60}$Co$_{20}$B$_{20}$ / Ta /  [Co(5 \r{A})/Pt(3 \r{A})]$_{\times 4}$  / Ru similar to  as in ref.~\onlinecite{swerts_BEOL_2015, devolder_evolution_2016} to form a bottom-pinned magnetic tunnel junction with properties designed for spin-torque applications \cite{devolder_time-resolved_2016}. 
All samples were annealed at 300$^{\circ}$C for 30 minutes in an out-of-plane field of 1 T. 

%%%%%%%%%%%%%%%%%%%%%%%%%%%%%%%%%%%%%%%%%%%%%%%%%%%%%%%%%%%%%%
%	Figure
%%%%%%%%%%%%%%%%%%%%%%%%%%%%%%%%%%%%%%%%%%%%%%%%%%%%%%%%%%%%%%

%%%%%%%%%%%%%%%%%%%%%%%%%%%%%%%%%%%%%%%%%%%%%%%%%%%%%%%%%%%%%%
\section{structure}
X-ray reflectivity scans (not shown) indicate Bragg reflexions at $2\theta=11$, 22.2 and 33.6 deg., consistent with the multilayer periodicity of 8 \r{A}. Consistently, the Pt to Co intermixing is sufficiently low that well formed 3\r{A} Pt spacers can be seen the Transmission Electron Micrograph after annealing [Fig.~\ref{TEM}(a)]. Almost no roughness is observed throughout the Co/Pt multilayer. We emphasize that this quality of interfaces is almost equivalent to that obtained in Molecular Beam Epitaxy conditions \cite{weller_growth_2001}. Indeed Co and Pt are strongly miscible such that hyperthermal (high energy) deposition techniques like sputter deposition do not easily yield this low degree of intermixing, except when the deposition is conducted under sufficiently low plasma pressure \textcolor{black}{in remote plasma conditions, i.e. when the substrate-to-target distance is large to avoid direct plasma exposure to the film being deposited}.

%%%%%%%%%%%%%%%%%%%%%%%%%%%%%%%%%%%%%%%%%%%%%%%%%%%%%%%%%%%%%%
\section{Anisotropy}
The magnetic material properties were measured by vibrating sample magnetometry (VSM) and Vector Network Analyzer ferromagnetic resonance \cite{bilzer_vector_2007} in both easy (z) and hard axis (x) configurations. \textcolor{black}{For VNA-FMR the sample is mechanically pressed on the surface of a 50 microns wide coplanar waveguide terminated by an open circuit; data analysis is conducted following the methods described in ref. \onlinecite{bilzer_open-circuit_2008}}. The VSM signal indicated a magnetization $M_s = 8.5\times10^{5}$ kA/m if assuming a magnetic thickness of 48 \r{A}, i.e. assuming that the [Co(5 \r{A})/Pt(3 \r{A})]$_{\times 6}$ multilayer can be described as a single material. 
The loops indicate a perpendicular anisotropy with full remanence. The reversal starts at 46.8 mT and completes before 48 mT with a tail-free square hysteresis loop. Careful attempts to demagnetize the sample using %either 
an \textit{ac} perpendicular field %or a saturating in-plane field both 
failed to produce a multidomain state at remanence. This indicates that the lowest nucleation field in the whole sample is larger that the domain wall propagation field everywhere in the film. This low propagation field indicates qualitatively that the effective anisotropy field is very uniform. The hard axis loop indicates an in-plane saturation field of $\approx1.3\pm0.1 ~\textrm{T}$ in line with the expectations for such composition \cite{chappert_planar_1998}. The rounding of the hard axis loop near saturation and its slight hysteretic remanence [Fig.~\ref{TEM}(c)] impedes a more precise deduction of the anisotropy fields from the sole hard axis loop. 
%%%%%%%%%%%%%%%%%%%%%%%%%%%%%%%%%%%%%%%%%%%%%%%%%%%%%%%%%%%%%%
%	Figure
%%%%%%%%%%%%%%%%%%%%%%%%%%%%%%%%%%%%%%%%%%%%%%%%%%%%%%%%%%%%%%
\begin{figure}
\includegraphics[width=8 cm]{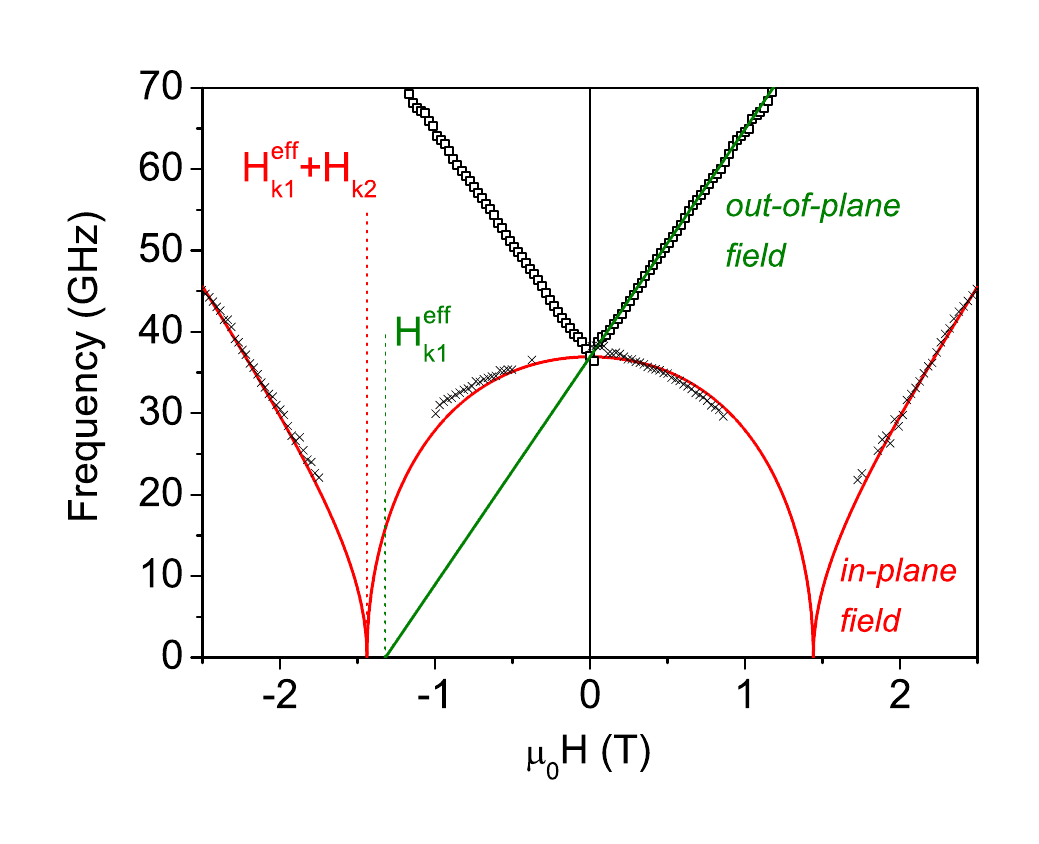}
\caption{(Color online). FMR frequencies versus in-plane (cross symbols) or out-of-plane (square symbols) applied field. The bold lines are fits using Eq.~\ref{omegaperp} and \ref{omegapara}, yielding $\mu_0(H_{k1}-M_s) = 1.320  \pm 0.005~\textrm{T}$ and $\mu_0H_{k2} = 0.120 \pm 0.015~\textrm{T}$.}
\label{FMR}
\end{figure}

We shall instead use the ferromagnetic resonance data because magnetization eigenfrequencies constitute absolute measurements of the effective fields acting on the magnetization. Fig. 2 gathers the measured FMR frequencies measured for in-plane and out-of-plane applied fields from -2.5 to 2.5 T. To analyze the microwave susceptibility data, we assume an energy density that reads $E =  \frac{1}{2} \mu_0 H_{k1} M_S ~\textrm{sin}^2\theta + \frac{1}{4} \mu_0 H_{k2} M_S ~\textrm{sin}^4\theta$ with $\theta$ the (supposedly uniform) angle between the magnetization and the sample normal. Our convention is that the first and second order magneto-crystalline anisotropy fields $H_{k1} = 2K_1/ (\mu_0 M_S)$ and $H_{k2} = 4K_2/ (\mu_0 M_S)$ are positive when they favor perpendicular magnetization, i.e. $\theta = 0$. 

In that framework, the ferromagnetic resonance frequencies in out-of-plane and in-plane applied fields saturating the magnetization as:
\begin{equation} \omega_{\textrm{perp}} = \gamma_0 (H_z + H_{k1}-M_s) \label {omegaperp} \end{equation} and
\begin{equation} \omega_{\textrm{in-plane}} = \gamma_0 \sqrt{H_x (H_x - H_{k1} - H_{k2} + M_s)}~, \label{omegapara}\end{equation}

where $\gamma_0=|\gamma | \mu_0$ is the gyromagnetic ratio. 
(For in-plane fields $H_x$ lower than $H_{x,~\textrm{sat}} = H_{k1} - H_{k2} -M_s$ the magnetization is tilted. A straightforward energy minimization was used to yield magnetization tilt $\theta$ that was subsequently injected to a Smit and Beljers equation to yield the FMR frequency). The best fit to the experimental data is obtained for $\mu_0(H_{k1}-M_s) = 1.320  \pm 0.005~\textrm{T}$ (corresponding to $K_1 = 10^6~\textrm{J/m}^3$) and $\mu_0H_{k2} = 0.120 \pm 0.015~\textrm{T}$. Note that the second order anisotropy is small but non negligible such that the effective anisotropy fields deduced from easy and axis axis measurements would differ by circa 10\% if $H_{k2}$ was disregarded.

%%%%%%%%%%%%%%%%%%%%%%%%%%%%%%%%%%%%%%%%%%%%%%%%%%%%%%%%%%%%%%
%%%%%%%%%%%%%%%%%%%%%%%%%%%%%%%%%%%%%%%%%%%%%%%%%%%%%%%%%%%%%%
%%%%%%%%%%%%%%%%%%%%%%%%%%%%%%%%%%%%%%%%%%%%%%%%%%%%%%%%%%%%%%
\section{Gilbert damping}
\subsection{Models}
We now turn to the analysis of the FMR linewidth (Fig. 3). As common in FMR, the linewidth comprises an intrinsic Gilbert damping part and an extrinsic additional contribution linked to the lateral non uniformity of the local effective fields $H_{k1} -M_s$. This can be gathered in a characteristic field $\Delta H_0$ measuring the disorder relevant for FMR. In out-of-plane field FMR experiments, the proportionality between effective fields and resonance frequencies (Eq. \ref{omegaperp}) allows to write simply $\Delta H_0 = \frac{1}{\gamma_0} \Delta \omega |_{\omega \rightarrow 0}$, and for the perpendicular magnetization we follow the usual convention \cite{shaw_determination_2012} and write: 
\begin{equation} 
\frac{1}{\gamma_0} \Delta \omega_{\textrm{perp}} =  2 \alpha (H_z + H_{k1}- M_s) + \Delta H_0  
\end{equation} 
or equivalently $\Delta \omega_{\textrm{perp}} =2 \alpha \omega_{\textrm{perp}} +  \gamma_0 \Delta H_0$.

For in-plane magnetization, the intrinsic linewidth above the in-plane saturation field is 
\begin{equation} \frac{1}{\gamma_0} \Delta \omega_{\textrm{in-plane}}^{\textrm{Gilbert}} =\alpha (2 H_x - H_{k1} - H_{k2} +M_s)
\label{linewidthHx} \end{equation}
The resonance frequency (Eq. \ref{omegapara}) is non linear with the effective fields such that the non uniformity $\Delta H_0$ of the local effective fields translates in a linewidth broadening through the term 
\begin{equation}
\frac{1}{\gamma_0} \Delta \omega_{\textrm{in-plane}}^{\textrm{disorder}} = \frac{d \omega_\textrm{in-plane}}{d {(M_s - H_{k1}})} \Delta H_0 
\label{linewidthHxextrinsic} \end{equation}
where the derivative term is $ \frac{\sqrt{H_x}}{2\sqrt{H_x - H_{k1} - H_{k2} +M_s}}$. In case of finite disorder, this factor diverges at the spatially-averaged in-plane saturation field $H_{x,~\textrm{sat}}$.

\subsection{Results} %%%%%%%%%%%%%%%%%%
For each applied field, the real and imaginary parts of the transverse permeability $\mu(f)$ were fitted with the one expected for the uniform precession mode \cite{devolder_using_2017} with three free parameters: the FMR frequency $\omega_{\textrm{FMR}} / (2 \pi)$, the FMR linewidth $\Delta \omega / (2 \pi))$ and a scaling (sensitivity) factor common to both real and imaginary parts of $\mu(f)$ as illustrated in Fig. 3b. 

When plotting the symmetric lorentzian-shaped imaginary part of the transverse permeability versus the asymetric lorentzian-shaped real part of the permeability for frequencies ranging from \textit{dc} to infinity, a circle of diameter $M_s / [2 \alpha (H_z+H_{k1}-M_s)]$ should be obtained for a spatially uniform sample \cite{liu_seed_2017}. The finite disorder $\Delta H_0$ distorts the experimental imaginary part of the permeability towards a larger and more gaussian shape. It can also damp and smoothen the positive and negative peaks of the real part of the permeability; when the applied field is such that the inhomogeneous broadening is larger than the intrinsic Gilbert linewidth, this results in a visible ellipticity of the polar plot of $\mu(f)$. In our experimental polar plot of $\mu(f)$ (Fig. 3a) the deviations from perfect circularity are hardly visible which indicates that the inhomogeneous broadening is not the dominant contribution to the sample FMR linewidth in out-of-plane field conditions.

To confirm this point we have plotted in Fig. 3c the dependence of FMR linewidth with FMR frequency for out-of-plane applied fields. A linear fit yields $\alpha = 0.021 \pm 0.002$ and $\Delta H_0 \approx 40~\textrm{mT}$. A substantial part of the measured linewidth thus still comes from the contribution of the lateral inhomogeneity of the effective anisotropy field within the film. As a result, low field measurements of the FMR linewidth would be insufficient to disentangle the Gilbert contribution and the structural disorder contributions to the total FMR linewidth.

The in-plane applied field FMR linewidth can in principle be used to confirm this estimate of the damping factor. Unfortunately we experience a weak signal to noise ratio in in-plane field FMR experiments such that only a crude estimation of the linewidth was possible.Within the error bar, it is independent from the applied field from 1.7 to 2.5 T (not shown) which indicates that the disorder still substantially contributes to the linewidth even at our maximum achievable field. At 2.5 T the linewidth was $\frac{1}{2 \pi}\Delta \omega_{\textrm{in-plane}} \approx 3.0\pm0.3 \textrm{~GHz}$. This is consistent width the expectations of that would predict 2.2 GHz of intrinsic contribution (Eq.~\ref{linewidthHx}) and 0.4 GHz of intrinsic contribution (Eq.~\ref{linewidthHxextrinsic}).

%%%%%%%%%%%%%%%%%%%%%%%%%%%%%%%%%%%%%%%%%%%%%%%%%%%%%%%%%%%%%%
%	Figure
%%%%%%%%%%%%%%%%%%%%%%%%%%%%%%%%%%%%%%%%%%%%%%%%%%%%%%%%%%%%%%
\begin{figure}
\includegraphics[width=8 cm]{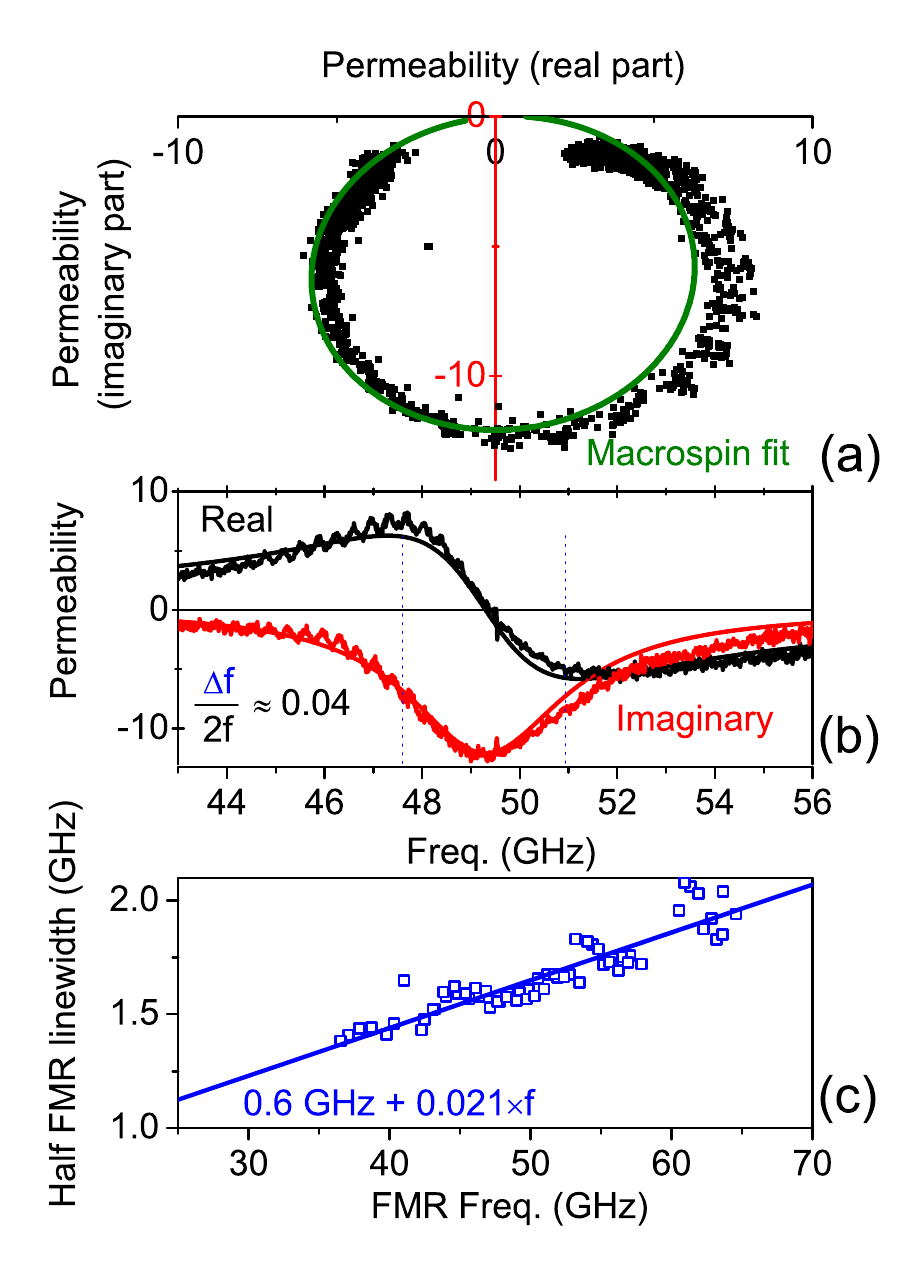}
\caption{(Color online). Gilbert damping of the Co/Pt multilayer. (a) Imaginary part versus real part of the permeability for a field of 0.45 T applied perpendicularly to the plane. The bold lines are theoretical macrospin permeability curves with linewidth parameters (i.e. effective damping) of 0.04. (b) Same data but versus frequency.  (c): FMR half linewidth versus FMR frequency. The bold line is a guide to the eye with a slope $\alpha=0.021$ and zero frequency intercept of 0.6 GHz.}
\label{damping}
\end{figure}

%%%%%%%%%%%%%%%%%%%%%%%%%%%%%%%%%%%%%%%%%%%%%%%%%%%%%%%%%%%%%%
\section{Discussion}

We conclude that the damping of Co/Pt multilayers can be of the order of 0.02 even for multilayers with anisotropies among the strongest reported (see ref. \onlinecite{guo_survey_2006} for a survey of the anisotropy of Co/Pt multilayers). Note that $\alpha \approx 0.021$ is still a higher bound, as we are unable to measure and subtract the spin-pumping contribution. Measuring the spin-pumping contribution would require to vary the cap and buffer layer thicknesses without affecting the multilayer structure which is difficult to achieve. Still, we can conclude that the damping of Co/Pt multilayers lies in the same range as other high anisotropy multilayers like Co/Ni (ref. \onlinecite{beaujour_ferromagnetic_2007, liu_seed_2017}) and Co/Pd (ref. \onlinecite{shaw_determination_2012}) systems. 

This conclusion is in stark contrast with the common thinking \cite{metaxas_creep_2007} that Co/Pt systems \textit{alway}s have a large damping. This widespread opinion is based on the standard models of magneto-crystalline anisotropy \cite{bruno_tight-binding_1989} and damping \cite{kambersky_spin-orbital_2007} that predicts that they both scale with the square of the spin-orbit coupling $\xi$, which is particularly large in the Pt atoms. We emphasize that this expectation of large damping is not systematically verified: in studies that make a thorough analysis of the effects of structural disorder, no correlation was found between anisotropy and damping in comparable material systems \cite{mizukami_gilbert_2010, shaw_determination_2012}. Rather, a large correlation was found between $H_{k1}$ and $\Delta H_0$, indicating that when the anisotropy is strong, any local inhomogeneity thereof has a large impact on the FMR linewidth.
Owing to the difficulty of achieving well-defined Co/Pt interfaces, we believe that past conclusions on the large damping of Co/Pt systems were based on systems likely to present some intermixing at the interface; indeed the presence of impurities with large spin-orbit coupling considerably degrades (increases) the damping of a magnetic material \cite{rantscher_effect_2007} and synchonously degrades (decreases) the magneto-crystalline anisotropy \cite{devolder_light_2000}.

%%%%%%%%%%%%%%%%%%%%%%%%%%%%%%%%%%%%%%%%%%%%%%%%%%%%%%%%%%%%%%
\section{Conclusion}
In summary, we have studied high anisotropy [Co(5 \r{A})/Pt(3 \r{A})]${\times 6}$ multilayers grown by low pressure remote plasma sputter deposition. The deposition conditions were tuned to achieve abrupt interfaces with little intermixing. Broadband ferromagnetic resonance was used to measure the first and second order uniaxial anisotropy fields. With the magnetization measured by vibrating sample magnetometry, this yields an anisotropy energy of $1~\textrm{MJ/m}^3$. 
The inhomogeneous broadening accounts for part of the ferromagnetic resonance linewidth, indicating some structural disorder leading to a equivalent 40 mT (or equivalently 600 MHz) of inhomogenity of the effective field in out-of-plane applied fields. This FMR-relevant inhomogeneity is comparable to the coercivity of 47 mT. Despite the large anisotropy a Gilbert damping parameter as low as 0.021$\pm$0.002 is obtained. This unexpectedly relatively low damping factor indicates that the presence of the Pt heavy metal within the multilayer can in some condition not be detrimental to the damping. We interpret our results and literature values by analyzing the consequences of Pt/Co intermixing: Pt impurities within a Cobalt layer reduce locally the interface anisotropy as they reduce the abruptness of the composition profile, but they also increase substantially the Gilbert damping. As a result, a large anisotropy together with a low damping can be obtained provided that intermixing is minimized at the Co/Pt interfaces.

%%%%%%%%%%%%%%%%%%%%%%%%%%%%%%%%%%%%%%%%%%%%%%%%%%%%%%%%%%%%%%
%%%%%%%%%%%%%%%%%%%
%%%%%%%%%%%%%%%%%%%	
%\bibliography{bib.bib}

\end{document}